# Model for the robust establishment of precise proportions in the early *Drosophila* embryo

Tinri Aegerter-Wilmsen*, Christof M. Aegerter†‡ & Ton Bisseling*

*Laboratory of Molecular Biology, Wageningen UR, Dreijenlaan 3, 6703HA Wageningen, The Netherlands.*

*† Department of Physics and Astronomy, Vrije Universiteit, De Boelelaan 1081, 1081HV Amsterdam, The Netherlands.*

*‡Present address: Fachbereich Physik, Universität Konstanz, P.O. Box 5560, 78457 Konstanz, Germany.*

**Corresponding author**: Prof. Dr. T. Bisseling, Laboratory of Molecular Biology, Wageningen UR, Dreijenlaan 3, 6703HA Wageningen, The Netherlands; tel: +31 31 748 3265; Fax: +31 31 748 3584; e-mail: ton.bisseling@wur.nl

**Abstract**

**During embryonic development, a spatial pattern is formed in which proportions are established precisely. As an early pattern formation step in *Drosophila* embryos, an anterior-posterior gradient of Bicoid (Bcd) induces *hunchback* (*hb*) expression (Driever et al. 1989; Tautz et al. 1988). In contrast to the Bcd gradient, the Hb profile includes information about the scale of the embryo. Furthermore, the resulting *hb* expression pattern shows a much lower embryo-to-embryo variability than the Bcd gradient (Houchmandzadeh et al. 2002). An additional graded posterior repressing activity could theoretically account for the observed scaling. However, we show that such a model cannot produce the observed precision in the Hb boundary, such that a fundamentally different mechanism**

**must be at work. We describe and simulate a model that can account for the observed precise generation of the scaled Hb profile in a highly robust manner. The proposed mechanism includes Staufen (Stau), an RNA binding protein that appears essential to precision scaling (Houchmandzadeh et al. 2002). In the model, Stau is released from both ends of the embryo and relocalises *hb* RNA by increasing its mobility. This leads to an effective transport of *hb* away from the respective Stau sources. The balance between these opposing effects then gives rise to scaling and precision. Considering the biological importance of robust precision scaling and the simplicity of the model, the same principle may be employed more often during development.**

1. Introduction

During embryo development, morphogen gradients confer positional information and thereby determine cell fate (Wolpert 1969). If the positions of regions of different cell fate along a body axis are solely determined by the interpretation of one morphogen gradient, their boundary positions cannot be adapted to variation in embryo size. With a bipolar gradient system in contrast, an adaptation to embryo size can be achieved (Wolpert 1969). The generation of the *hb* expression domain along the anterio-posterior body axis, early during Drosophila development, seems to involve such a bipolar system. Bcd forms an anterior gradient and induces *hb* expression (Driever et al. 1989, Tautz et al. 1988, Struhl et al. 1989), whereas Nanos (Nos) forms a posterior gradient and blocks *hb* translation (Irish et al. 1989). As expected based on a bipolar system, the position of the Hb boundary, $x_{Hb}$, defined as the position where $[Hb](x_{Hb}) = 0.5\,[Hb]_{max}$, showed a strict linear correlation with embryo length (EL). In contrast, the position of the Bcd boundary, defined such that on average $x_{Hb} = x_{Bcd}$, is independent of EL (Houchmandzadeh et al. 2002). However, this scaling of the Hb profile is already established in the *hb* mRNA profile (Houchmandzadeh et al. 2002), which is not consistent with the above model in which Nos regulates *hb* mRNA translation. Furthermore, the Hb scaling property is not abolished in *nos* knock-outs (Houchmandzadeh et al. 2002), which further undermines a role for Nos in scaling the Hb profile. Theoretically, it is still conceivable that there is another, yet unknown, graded posterior repressing activity instead of Nanos.

However, not only scaling is important in the Hb boundary, but also precision (Houchmandzadeh et al. 2002). This is in line with earlier observations that changes in Bcd concentration induce only a relatively small shift in the positions of anterior markers (Driever et al. 1988). In fact, we will show in section 2 below that the observed precision of the Hb boundary (Houchmandzadeh et al. 2002) cannot be explained by such a bipolar gradient model from observations of the fluctuations of the Bcd gradient available online.

Thus, a fundamentally different model is required to explain the observed high precision of $x_{Hb}$ positioning. Furthermore, a model which describes the generation of the Hb profile should also account for the observed remarkable robustness of this process against temperature differences and it should evidently still account for the observed scaling. In order to identify genes that contribute to the scaling and precision of the Hb boundary, the Bcd and Hb profiles have been analysed in a range of different mutants (Houchmandzadeh et al. 2002). These include embryos with mutations in genes whose protein products are known to interact directly with Hb as well as embryos in which large chromosome parts were removed. Scaling and precision of the Hb profile were conserved in all mutants, except in *stau* mutants (Houchmandzadeh et al. 2002). Stau is a double stranded RNA binding protein, which plays an essential role in the localisation of several mRNAs during development (Broadus et al. 1997, Clark et al. 1994, Ferrandon et al. 1994, Li et al. 1997, Pokrywka et al. 1991, St. Johnston et al. 1992). It is located at the anterior as well as at the posterior end of an egg (St. Johnston et al. 1991), prior to the generation of the Hb profile. It remains a challenge to reveal the mechanism by which Stau can robustly induce precision and scaling in the Hb profile. In section 3, we formulate a model incorporating Stau for the process of Hb boundary formation with a set of reaction-diffusion equations. The model is numerically simulated and reproduces the precision and scaling observed in (Houchmandzadeh et al. 2002).

In section 4, we test the model's robustness to changes in the parameter values in terms of its capability to still produce a precise Hb boundary. In this context, we also test the effects of removing each of the Stau sources. Furthermore, the model is extended to include other factors that are known to influence the Hb expression pattern (Houchmandzadeh et al. 2002).

Finally, in section 5 we discuss the essential assumptions underlying the model, which can serve as experimentally testable hypotheses. Furthermore, due to the small number of parameters necessary to obtain the main features of precision and robustness, we argue that similar mechanisms may also play in other developmental systems.

## 2. Properties of the bipolar gradient model

In order to study quantitatively the properties of a possible bipolar gradient model, we consider a simplified model of pattern formation in the embryo using two competing gradients. Here, a Bcd gradient (promoting the expression of *hb*) is present anteriorly and a gradient is present posteriorly of a putative protein, pp, that inhibits *hb* expression. The promoting / inhibiting efficiency times the concentration of protein will thus determine the effective influence on *hb* expression, such that the point where *hb* changes from being expressed to being suppressed (the *hb* boundary $x_{hb}$) is given by the point where the effective influences of the two proteins are equal. As was experimentally seen, Bcd forms an exponential gradient (Houchmandzadeh et al. 2002). This is also expected from a gradient formation mechanism, which consists of a source at the edge and combined diffusion and break-down of the protein in the embryo. The characteristic fall-off of the exponential gradient is then given by $\lambda = (k/D)^{1/2}$, where k is the break-down rate and D is the Diffusion coefficient. We therefore assume that pp likewise forms an exponential gradient according to the same mechanism. Any changes in the gradients are then due to differences in the local environment (e.g. temperature differences), determining the viscosity of the intracellular plasma and thus the diffusion coefficient or the concentration/activity of protease and thus the break-down rate.

The position of the *hb* boundary can then be obtained from equating two exponential gradients corresponding to the Bcd and pp gradients:

$$K_{Bcd} \exp(-\lambda_{Bcd} x) = K_{pp} \exp(-\lambda_{pp}(EL-x)), \tag{1}$$

where $K_{Bcd}$ and $K_{pp}$ denote the effective influence of Bcd and pp respectively, $\lambda_{Bcd}$ and $\lambda_{pp}$ are the respective decay lengths and EL is the embryo length. Solving for x (the point of equal suppression and expression) one obtains:

$$x_{hb} = (\lambda_{pp} EL + \ln(K_{Bcd}/K_{pp}))/(\lambda_{Bcd} + \lambda_{pp}) = EL/(1+A) + \ln(B)/\lambda(1+A), \tag{2}$$

where $A = \lambda_{Bcd}/\lambda_{pp}$ and $B = K_{Bcd}/K_{pp}$. Supposing that the effective influence of Bcd and that of pp at the anterior and posterior end of the embryo respectively are on average the same, i.e. $<B> = 1$, the second term vanishes and perfect scaling (i.e. $x_{hb} \propto EL$) is obtained. This shows that an equal effective influence of both proteins is necessary in order to allow for scaling as found experimentally. Furthermore, in case the fall-off of Bcd and that of pp are comparable (i.e. $\lambda_{Bcd} \cong \lambda_{pp}$ or $<A> = 1$) this results in $x_{hb} = EL/2$, which is the scaling which was observed experimentally in wild type embryos as well as in mutant embryos in which scaling and precision was conserved (Houchmandzadeh et al. 2002).

Given this dependence of $x_{hb}$ on the characteristic length scales of the two proteins, we now estimate the maximally possible level of precision predicted by this model constrained by available experimental data. There are two possibilities: (i) that the variations in Bcd and pp are uncorrelated and (ii) that the opposing gradients are correlated and that thus we only have to consider fluctuations in the ratio of the gradients, A. In the first case, a lower bound for the variability, or error, in $x_{hb}$ is obtained from assuming that only $\lambda_{Bcd}$ fluctuates and $\lambda_{pp}$ does not vary. This results in:

$$\delta x_{hb} = |\partial x_{hb} / \partial \lambda_{Bcd}| \delta \lambda_{Bcd} = \lambda_{pp} EL / (\lambda_{Bcd} + \lambda_{pp})^2 \delta \lambda_{Bcd} = x_{hb} / (\lambda_{Bcd} + \lambda_{pp}) \delta \lambda_{Bcd}. \quad (3)$$

Again supposing that $\lambda_{Bcd} \cong \lambda_{pp}$, this simplifies to

$$\delta x_{hb} = x_{hb} / 2 * \delta \lambda_{Bcd} / \lambda_{Bcd}. \quad (4)$$

Experimentally, the relative error in $\lambda_{Bcd}$ was determined in (Houchmandzadeh et al. 2002), where it is stated that $\delta \lambda_{Bcd} / \lambda_{Bcd} = 0.2$. Thus using $x_{hb} = EL/2$ from above, we obtain $\delta x_{hb} = 0.05$ EL. This is a factor of five worse than what was experimentally observed in (Houchmandzadeh et al. 2002) ), implying that a bi-gradient model cannot produce the necessary precision in case the fluctuations of the two characteristic length scales are uncorrelated. For the second case, where they are correlated, the variability of $x_{hb}$ is determined by $\delta A$ via:

$$\delta x_{hb} = |\partial x_{hb} / \partial A| \delta A = \delta A\, EL / (1 + A)^2 = \delta A\, EL / 4. \quad (5)$$

Thus we have to estimate $\delta(\lambda_{Bcd} / \lambda_{pp})$. From the derivation of the expression for $x_{hb}$ (Eq. 2), it can be seen that in its determination only the anterior part of the Bcd gradient and the posterior part of the pp gradient are important, as only these respective parts of the gradients determine the point where their effective influences are equal. Thus one can write $\delta A = \delta(\lambda_{Bcd}^{ant} / \lambda_{pp}^{post})$. Now we assume that the two characteristic length scales vary in the same way, i.e. that they are perfectly correlated. Thus in this case, any local cause of the variations in the Bcd gradient is also affecting the pp gradient in the same way. Therefore, the posterior part of the pp gradient, $\lambda_{pp}^{post}$, can be estimated from the posterior part of the Bcd gradient, $\lambda_{Bcd}^{post}$. This leads to a lower bound for $\delta A > \delta(\lambda_{Bcd}^{ant} / \lambda_{Bcd}^{post})$, which is available experimentally. We have used the database of publicly available measurements of the Bcd gradient in various developmental stages of Drosophila at http://flyex.ams.sunysb.edu/flyex/ (Kozlov et al. 2000, Myasnikova et al. 2001). There, we have used 111 embryos from cleavage cycle 14A at time periods 1,2, and 3 to correspond to the experiments of (Houchmandzadeh et al. 2002). The Bcd concentration profile was then determined from an average in the central 10% of the embryo along the dorsal-ventral axis. Moreover, the background illumination intensity was subtracted in order to obtain exponential gradients, such that the variations of the logarithm of the intensities from a linear dependence were minimised. From these different gradients we subsequently determined the exponential decay constant, $\lambda_{Bcd}$, in the second and third quarter of the embryo, $\lambda_{Bcd}^{ant}$ and $\lambda_{Bcd}^{post}$. This was done because in the first quarter there are non-exponential deviations due to the peak at a finite length and in the last quarter there may be systematic deviations due to the background-intensity subtraction. The results below are however not changed by varying the position and length of the fitting window within 0.05 EL or by extending the portion of the embryo studied along the dorsal-ventral axis up to 20%. The ratio of the two thus determined characteristic scales, $\lambda_{Bcd}^{ant} / \lambda_{Bcd}^{post}$ is then calculated for each embryo as a measure of the parameter A in the model. Averaged over the 111 embryos available in the relevant time period, this ratio is $<\lambda_{Bcd}^{ant} / \lambda_{Bcd}^{post}> = 1.04$ with a standard deviation of $\delta(\lambda_{Bcd}^{ant} / \lambda_{Bcd}^{post}) = 0.18$. Thus we can conclude that $\delta A > 0.18$, which directly yields lower bound for $\delta x_{hb}$ = 0.045 EL using Eq. 5. Again, this is more than a factor of four worse than what is experimentally observed. Thus also if the variations of the gradients are perfectly correlated, a bigradient model cannot produce the required precision in the scaling of the hb boundary and must thus be missing an important biological ingredient.

In addition, the ratio of effective influence of the two proteins, the second term in Eq. 2, would most probably also have an error, $\delta B$, which gives a further contribution to the variability in $x_{hb}$, given by $\delta x_{hb} = \delta B / 2\lambda$.

## 3. Model for the generation of precise and scaled Hb boundary

In the model proposed here, an anterior Bcd gradient as well as Stau located anteriorly and posteriorly are initially present. Subsequently, the model not only allows for the known *hb* induction by Bcd (Driever et al. 1989, Tautz et al. 1988, Struhl et al. 1989), but it also assumes that Stau is gradually released from both ends of the embryo and can reversibly form a complex with *hb* mRNA. Furthermore, the resulting protein-mRNA complex has a higher mobility in the model than unbound *hb* mRNA. This leads to an effective transport of *hb* away from the respective Stau sources. On the anterior side, the *hb* production is higher due to the anterior presence of Bcd. The balanced opposing effects of both Stau sources in concert with this asymmetric production of hb then give rise to scaling in the middle of the embryo. The effective transport of hb mRNA is self regulated by the fact that the ensuing changes in the *hb* profile lead to changes in diffusive flux, which counteract the transport via Stau. This gives an effective control of the extent of Stau assisted transport and thus yields precision and robustness of the *hb* boundary.

In a mathematical terms, this implies that the initial conditions of the model consist of two sources of Stau ($[Stau]_{ini-ant}$ at the anterior side (x=0) and $[Stau]_{ini-post}$ at the posterior side (x=EL)) and an anterior Bcd gradient. This gradient was generated by numerically simulating the following reaction-diffusion equation until a stable state was reached. For all numerical simulations, LabView was used.

$$\partial[Bcd]/\partial t = D_{Bcd}\nabla^2[Bcd] + k_{Bcd\ tl}\ (0.03EL<x<0.13EL) - k_{Bcd\ br}[Bcd] \tag{6}$$

The parameters include the diffusion coefficients of Bcd ($D_{Bcd}$), Bcd production by translation ($k_{Bcd\ tl}$), and an aspecific breakdown of Bcd ($k_{Bcd\ br}$). The equations were solved in the region $0 < x < EL$. In order to obtain a variable input set such as observed *in vivo*, $D_{Bcd}$ was varied between 18 µm$^2$/s and 90 µm$^2$/s and EL was varied independently between 400 µm and 500 µm. Other parameters used to generate the complete input: $k_{Bcd\ transl} = 0.0015$ s$^{-1}$, $k_{Bcd\ break} = 0.003$ s$^{-1}$, $[Stau]_{ini-ant} = 100$, $[Stau]_{ini-post} = 450$. This input was used for each simulation unless stated differently. The Bcd gradients obtained can be seen in Fig. 1a and have similar characteristics as the set of Bcd profiles that was experimentally observed: the position of the Bcd boundary, $x_{Bcd}$, is not correlated with embryo length (Spearman's rank correlation, $r_s = 0.12$, $P = 0.3$; Fig. 1b) and the Bcd boundary positions have a standard deviation of $\sigma = 0.07$ EL, which is equal to the one measured experimentally (Houchmandzadeh et al. 2002).

Using these Bcd gradients as input, the generation of the Hb gradient is modelled by the following set of reaction-diffusion equations:

$$\partial[Stau]/\partial t = D_{Stau}\nabla^2[Stau] - k_{Stau\text{-}hb\ form}[Stau][hb] + k_{Stau\text{-}hb\ disint}[Stau\text{-}hb] - k_{Stau\ br}[Stau] + k_{Stau\text{-}ant\ rel}[Stau_{ant}](x=0) + k_{Stau\text{-}post\ rel}[Stau_{post}](x=EL) \tag{7}$$

$$\partial[Stau_{ant}]/\partial t = -k_{Stau\text{-}ant\ rel}[Stau_{ant}](x=0) \tag{8}$$

$$\partial[Stau_{post}]/\partial t = -k_{Stau\text{-}post\ rel}[Stau_{post}](x=EL) \tag{9}$$

$$\partial[hb]/\partial t = D_{hb}\nabla^2[hb] - k_{Stau\text{-}hb\ form}[Stau][hb] + k_{Stau\text{-}hb\ disint}[Stau\text{-}hb] - k_{hb\ br}[hb] + k_{tcr}[Bcd] \tag{10}$$

$$\partial[\text{Stau-}hb]/\partial t = D_{\text{Stau-}hb}\nabla^2[\text{Stau-}hb] + k_{\text{Stau-}hb\text{ form}}[\text{Stau}][hb] - k_{\text{Stau-}hb\text{ disint}}[\text{Stau-}hb] \quad (11)$$

$$\partial[\text{Hb}]/\partial t = k_{tl}[hb] \quad (12)$$

Here [Stau$_{ant}$] and [Stau$_{post}$] are the concentrations of Stau at the anterior and the posterior end respectively; D$_{Stau}$, D$_{hb}$ and D$_{Stau-hb}$ are the diffusion coefficients of Stau, *hb* and a complex of Stau protein and *hb* RNA respectively; k$_{Stau-hb\text{ form}}$ is the formation constant of the Stau-*hb* complex; k$_{Stau-hb\text{ disint}}$ is the disintegration constant of the Stau-*hb* complex; k$_{Stau\text{ br}}$ and k$_{hb\text{ br}}$ are constants for aspecific Stau and *hb* RNA breakdown respectively; k$_{Stau-ant\text{ rel}}$ and k$_{Stau-post\text{ rel}}$ are constants for the gradual release of Stau from the anterior and the posterior end respectively; k$_{tcr}$ is a constant for the transcription regulation of *hb* by Bcd and k$_{tl}$ is a constant for *hb* translation. These equations were solved in the region 0 < x < EL until [Stau]$_{total}$ < 0.05 [Stau]$_{total}$(t=0). For EL, the same values were chosen as for the generation of the input. In this model, *hb* transcription is linearly regulated by Bcd and *hb* translation depends linearly on the *hb* concentration. Using a tangent-hyperbolic dependence, which yields a biologically more realistic S-function, does not change the results significantly (data not shown). Moreover, Stau-*hb* movement was simulated as diffusion. It is however possible to construct a model in which anterior Stau can transport *hb* in a posterior direction only and vice versa, again with similar results (data not shown).

The following parameters were used to model the wild type situation: D$_{Stau}$ = 40μm$^2$/s (comparable with GFP diffusion constant in cytoplasm (Yokoe et al. 1996)), D$_{hb}$ = 8μm$^2$/s, D$_{Stau-hb}$ = 40μm$^2$/s, k$_{Stau-hb\text{ form}}$ = 3.5 s$^{-1}$, k$_{Stau-hb\text{ disint}}$ = 0.015 s$^{-1}$, k$_{Stau\text{ br}}$ = 0.015 s$^{-1}$, k$_{hb\text{ br}}$ = 0.015 s$^{-1}$, k$_{Stau-ant\text{ rel}}$ = 0.05 s$^{-1}$, k$_{Stau-post\text{ rel}}$ = 0.05 s$^{-1}$, k$_{tcr}$ = 0.005 s$^{-1}$, k$_{tl}$ = 0.00015 s$^{-1}$. In all of the above, concentrations are normalised to the maximum value of [Bcd].

To assess whether this model is indeed capable of generating a scaled Hb boundary in a highly precise manner when faced with variable Bcd gradients such as they were found among different embryos, the model was simulated with a set of different Bcd gradients (Fig. 1a). Provided that Stau was released from both ends of the embryo, it was possible to choose the largely unknown values of the different parameters in the model in such a way (see above) that simulations with the input set of variable Bcd gradients generated the Hb boundary which shows the required precision and scaling with $x_{Hb}$ at the *in vivo* position, 0.49 EL (Fig. 1c+d). The variability of the Hb boundary position $x_{Hb}$, which was generated with the simulations ($\sigma = 0.009$ EL), is similar to the experimentally observed one ($\sigma = 0.010$ EL). The Spearman's rank correlation coefficient, $r_s$, between $x_{Hb}$ and EL is 0.94, which is even higher than the experimentally obtained value ($r_s=0.82$), showing that the correlation between EL and $x_{Hb}$ is even stronger than measured experimentally. This correlation is already present at *hb* mRNA level, which is consistent with the experimental data. Thus, the model can account for the positioning of the Hb boundary in at least as precise and scaled a manner as was observed in the wild type embryo.

## 4. Robustness of the model and further extensions

It was also shown that the positioning of the Hb boundary is robust against considerable changes in temperature (Houchmandzadeh et al. 2002). This implies that any model describing the process has to be robust against changes in the values of its parameters as these are generally temperature dependent. To check for robustness of the model above described, each parameter in the model was halved and doubled respectively. Both the average position, its variance and scaling remain within tight limits, $0.475$ EL $< x_{Hb} < 0.505$ EL, $\sigma < 0.016$ EL and $r_s > 0.85$ (Fig. 2), showing that the model indeed generates the precisely scaled Hb boundary in a very robust manner. However, if Stau is absent at the ends of the embryo, the model does not result in a precise and scaled Hb boundary. When Stau is present only posteriorly, $x_{Hb} = 0.28$ with $\sigma = 0.03$ and $r_s = 0.5$; similarly, if Stau is present only anteriorly, $x_{Hb} = 0.51$, $\sigma = 0.11$ and $r_s = -0.5$. This is in line with observations of Stau mutants, where in the absence of Stau, precision and scaling were destroyed completely (Houchmandzadeh et al. 2002). A direct comparison with the Stau mutants is difficult however due to the multiple functions of Stau in the developmental process.

*In vivo*, the formation of the Hb gradient is mediated not only by Bcd and Stau, but also by other factors. For instance, it is known that Hb auto-induction occurs. Furthermore, there is initially equally distributed maternal *hb* RNA present and a posterior-anterior gradient of Nos blocks the translation of this maternal *hb* mRNA, such that, effectively, an anterior-posterior gradient of maternal *hb* RNA is initially present. Extending the model with these factors should evidently not lead to a disturbed precision, scaling and robustness, as such an extended model mimics the wild type situation better and in the wild type situation this precise and robust scaling is present.

In an extension of the model, we took these factors into account by assuming that the initial *hb* gradient has a similar form as the Bcd gradient and have modelled this *hb* gradient as being proportional to the initial Bcd gradient: $[hb]_{initial} = \text{mat}_{hb} [Bcd]$. Moreover, the auto-induction of *hb* in the zygote was simulated by adding a corresponding term to Eq. 10 describing the temporal variation of [*hb*] proportional to [Hb] yielding

$$\partial[hb]/\partial t = D_{hb}\nabla^2[hb] - k_{\text{Stau-}hb\text{ form}}[\text{Stau}][hb] + k_{\text{Stau-}hb\text{ disint}}[\text{Stau-}hb] - k_{hb\text{ br}}[hb] + k_{\text{tcr}}[Bcd] + k_{\text{auto}}[Hb]. \tag{10'}$$

The parameters chosen for the values of $\text{mat}_{hb}$ and $k_{\text{auto}}$ were such that in combination they led to an increase of Hb present at the end of the simulation, which is a factor of 4 higher than in the original simulation with the version of the model described in section 2. The contribution of auto-induction and maternal RNA to the total amount of Hb was varied, since they are not precisely known to our knowledge. The specific combinations were $mat_{hb} = 0$; $k_{auto} = 0.0015$ s$^{-1}$; $mat_{hb} = 0.1$; $k_{auto} = 0.001$ s$^{-1}$; $mat_{hb} = 0.3$; $k_{auto} = 0.0005$ s$^{-1}$; $mat_{hb} = 0.4$; $k_{auto} = 0.00015$ s$^{-1}$; $mat_{hb} = 0.5$; $k_{auto} = 0$ s$^{-1}$. For all of these combinations, precision and scaling were conserved, as indicated by the fact that $\sigma_{Hb} < 0.011$ EL and $r_s > 0.90$ for all of the above cases. Dependent on the precise contribution of both factors, the position of $x_{Hb}$ varied from 0.470 EL to 0.495 EL. Such small shifts in $x_{Hb}$ are to be expected based on the observation that $x_{Hb}$ varies from 0.43 EL to 0.54 EL among several mutants in which precision and scaling are conserved (Houchmandzadeh et al. 2002). Furthermore, a complete test of robustness was carried out, such as was done before in section 3, for all parameter combinations above. This involved the halving and doubling of every parameter in the model. For all of these simulations, $r_s$, $\sigma_{Hb}$, as well as the variation of $x_{Hb}$ around the value, $x^0_{Hb}$, obtained with the original parameter set, stayed within tight limits (i.e. $r_s > 0.80$, $\sigma_{Hb} < 0.018$, and $x^0_{Hb} - 0.015 < x_{Hb} < x^0_{Hb} + 0.015$).

Thus, in this extended model, which includes the auto-induction of Hb and an initially present anterior *hb* gradient, the precision, scaling and robustness are not disturbed from the minimal model, even when the total Hb amount was quadrupled. This was irrespective of the contribution of each factor to the total amount of Hb.

**5. Conclusions**

In conclusion, we have shown that a bipolar gradient model, such as is usually used to explain scaling during development, cannot account for the observed precision in the position of the Hb boundary. We have then presented an alternative model, which does produce a Hb boundary to the required precision in a robust manner. For the model, it is necessary that there is a source at each end of the embryo from which Stau is released. In a freshly laid egg, Stau is concentrated anteriorly as well as posteriorly (St. Johnston et al. 1991). For posteriorly localised Stau it was suggested that it gradually diffuses away during the syncytial blastoderm stages (St. Johnston et al. 1991), including those during which the Hb profile is formed. We predict that Stau is also released from the anterior end, but this remains to be studied. We also predict that Stau can form a complex with *hb* mRNA and thereby increase the mobility of *hb* mRNA. As was mentioned before, in the specific parameter set, which was used in this paper, the diffusion rate of the complex of Stau and *hb* mRNA was comparable to that of GFP in cytoplasm (Yokoe et al. 1996) and the diffusion rate of free *hb* mRNA was five times smaller. As can be concluded from the robustness test, the precise absolute values as well as the precise difference in diffusion rate is not crucial. It was even possible to use a model with directed transport instead of diffusion. Therefore, the model allows for different molecular mechanisms to increase the mobility of *hb* mRNA. Even though a role for Stau in Hb localisation is strongly suggested by the disturbed Hb localisation in certain Stau mutants, further studies are needed to confirm a direct involvement of Stau in *hb* mRNA localisation and to clarify the underlying mechanism. Stau has been shown to be involved in the localisation of other mRNAs. This includes the microtubuli dependent localisation of *oskar* (*osk*) mRNA (Clark et al. 1994) and *bcd* mRNA (Ferrandon et al. 1994, Pokrywka et al. 1991) as well as the actin dependent *prospero* (*pros*) mRNA localisation (Broadus et al. 1997, Li et al. 1997). A certain *stau* mutant (*stau$^{D3}$*) shows a severely disturbed localisation of each of these three mRNA species (Ferrandon et al. 1994, Broadus et al. 1998). However, the Hb gradient still shows relatively low variability (Houchmandzadeh et al. 2002). This suggests that, if Stau is indeed directly involved in localising *hb* mRNA, this occurs differently than for *osk*, *bcd* and *pros* mRNA.

The potential of differential mobility has previously been argued to be a powerful means to generate patterning (Koch et al. 1994, Turing 1952). Here we show that it can effectively create scaling, precision and robustness in a pattern initially generated by a morphogen gradient. The core of the proposed mechanism only consists of very few elements with few interactions among them. In combination with the biological relevance of robust precision scaling, this opens the possibility that the same principle may also be employed during other developmental patterning processes.

**Acknowledgements**

We would like to thank Ernst Hafen for his comments on the manuscript.

Yokoe, H. & Meyer, T. 1996. Spatial dynamics of GFP-tagged proteins investigated by local fluorescence enhancement. *Nat Biotechnol* **14**, 1252-6.

**Figure Legends**

Figure 1. The model yields a precise and scaled Hb profile when faced with variable and unscaled Bcd gradients. a. A set of 100 Bcd gradients that was generated by varying the Bcd diffusion coefficient and the embryo length independently. c. Hb profiles that were obtained with the Bcd input set that is shown in a. b, d. Position at which the Bcd gradient (b) and the Hb profile (d) cross a threshold concentration of 0.19 and 0.50 respectively versus embryo length.

Figure 2. The model can generate a precise and scaled Hb profile in a highly robust manner. a: the variation in the mean of $x_{Hb}$, b: the variation of $\sigma$, c: the variation of $r_s$. Horizontally, the different parameters have first been doubled and then halved in the order of $k_{Stau\ br}$, $k_{Stau-ant\ rel}$, $k_{Stau-post\ rel}$, $D_{Stau}$, $[Stau]_{ini}$, $k_{hb\ br}$, $k_{tcr}$, $D_{hb}$, $k_{tl}$, $D_{Stau-hb}$, $k_{Stau-hb\ form}$, $k_{Stau-hb\ disint}$, corresponding to increasing numbers 0 through 23.

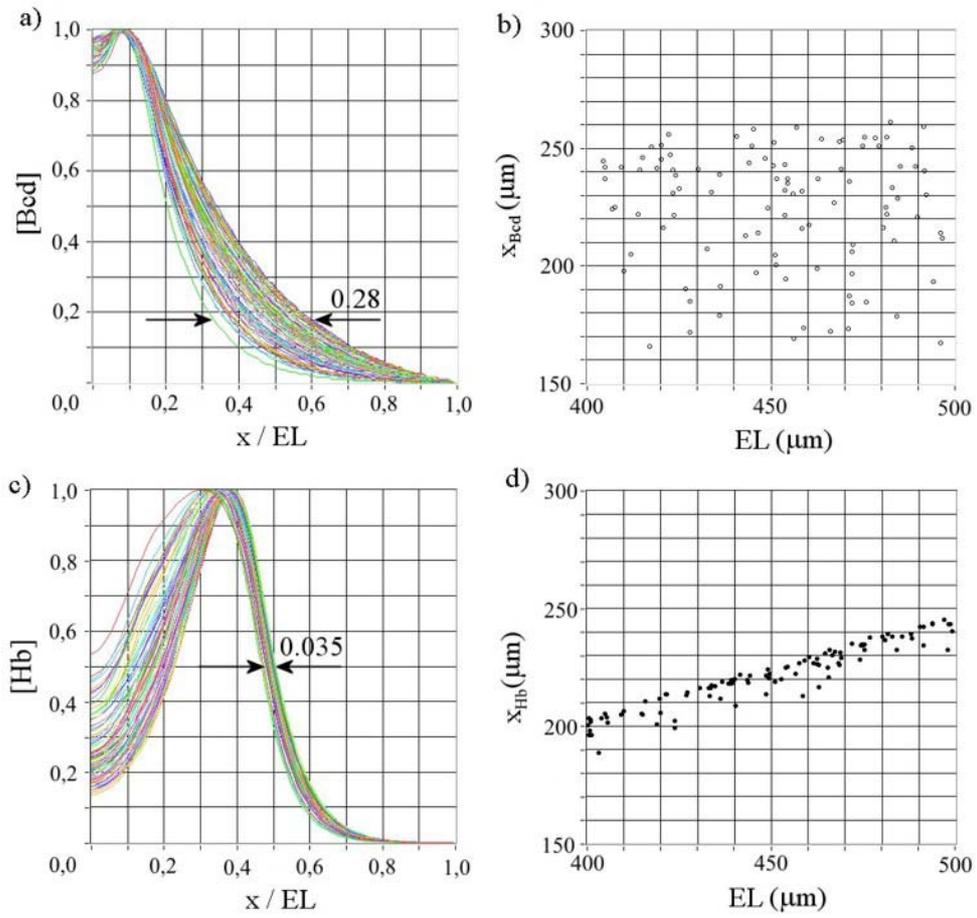

Figure 1

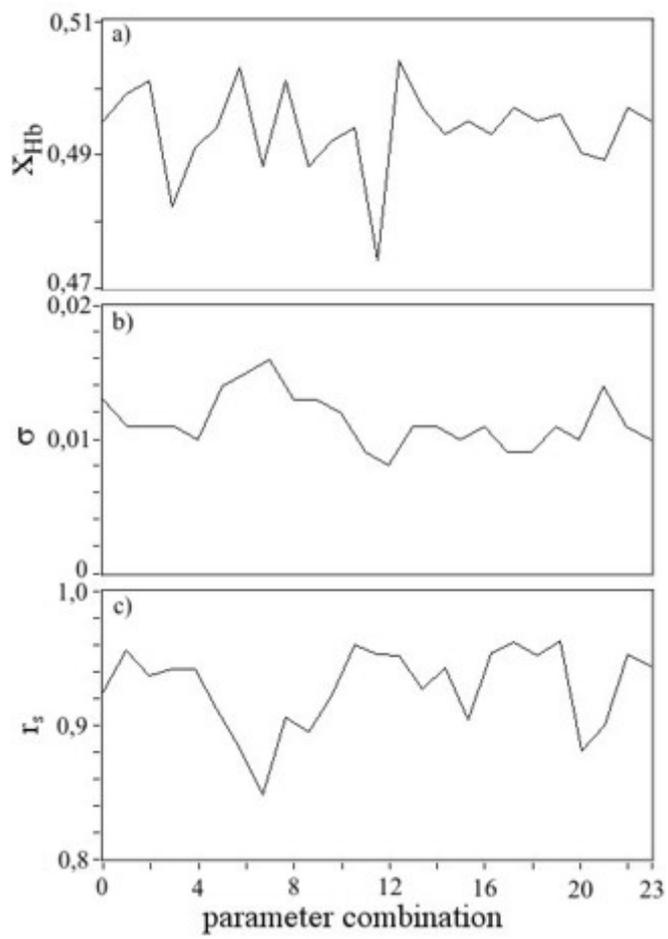

Figure 2